\documentclass[reprint,amsmath,amssymb,floatfix, twocolumn]{revtex4-1}

\usepackage{multirow}
\usepackage{bbold}
\usepackage{graphicx}
\usepackage{color}
\usepackage{hyperref}

\usepackage[caption=false]{subfig}
\usepackage[T1]{fontenc}

\usepackage[normalem]{ulem}

\usepackage{physics}
\usepackage{mathtools}
\usepackage{bbm}

\usepackage{amstext}
\usepackage{array}

\usepackage{diagbox}

\usepackage{siunitx}

\usepackage{yfonts}

\allowdisplaybreaks

\usepackage{setspace}

\newcommand{\ncmd}{\newcommand}
\ncmd{\nn}{\nonumber}
\ncmd{\pg}[1]{\textcolor{red}{#1}}
\ncmd{\mbf}[1]{\bs{#1}}
\ncmd{\Lam}{\Lambda}
\ncmd{\lam}{\lambda}
\ncmd{\Gam}{\Gamma}
\ncmd{\gam}{\gamma}
\ncmd{\sig}{\sigma}
\ncmd{\Dl}{\Delta}
\ncmd{\dl}{\delta}
\ncmd{\kap}{\kappa}
\ncmd{\Om}{\Omega}
\ncmd{\om}{\omega}
\ncmd{\mc}{\mathcal}
\ncmd{\eps}{\epsilon}
\ncmd{\veps}{\varepsilon}
\ncmd{\vphi}{\varphi}
\ncmd{\vtheta}{\vartheta}
\ncmd{\note}[1]{{\color{red}{#1}}}
\ncmd{\new}[1]{{\texttt{#1}  } }
\ncmd{\eq}[1]{Eq. \eqref{#1}}
\ncmd{\bs}{\boldsymbol}
\ncmd{\pll}{\parallel}
\ncmd{\dsty}{\displaystyle}
\ncmd{\methods}{\note{`Materials and Methods'}}
\ncmd{\suppl}{\note{`Supplementary Information'}}

\begin{document}
\title{Mixed-order topology of Benalcazar-Bernevig-Hughes models}
\author{Shouvik Sur$^{1,2}$}
\thanks{These authors contributed equally.}
\author{Alexander C. Tyner$^{3}$}
\thanks{These authors contributed equally.}
\author{Pallab Goswami$^{1,3}$}
\affiliation{$^{1}$ Department of Physics and Astronomy, Northwestern University, Evanston, IL 60208}
\affiliation{$^{2}$ Department of Physics and Astronomy, Rice University, Houston, Texas 77005, USA}
\affiliation{$^{3}$ Graduate Program in Applied Physics, Northwestern University, Evanston, IL 60208}
\date{\today}

\begin{abstract}
Benalcazar-Bernevig-Hughes (BBH) models, defined on $D$-dimensional simple cubic lattice, are paradigmatic toy models for studying $D$-th order topology and corner-localized, mid-gap states. Under periodic boundary conditions, the Wilson loops of non-Abelian Berry connection of BBH models along all high-symmetry axes have been argued to exhibit gapped spectra, which predict gapped surface-states under open boundary conditions. 
In this work, we identify 1D, 2D, and 3D topological invariants for characterizing higher order topological insulators. 
Further, we demonstrate the existence of cubic-symmetry-protected, gapless spectra of Wilson loops and surface-states along the body diagonal directions of the Brillouin zone of BBH models. 
We show the gapless surface-states are described by $2^{D-1}$-component, massless Dirac fermions. Thus, BBH models can exhibit the signatures of first and $D$-th order topological insulators, depending on the details of externally imposed boundary conditions.
\end{abstract}

\maketitle

\section{Introduction} \label{Intro} In quest for quantized, higher-order, electric multipole moments in crystalline systems, Benalcazar \emph{et al.} introduced the intriguing concept of higher-order topological insulators (HOTI) \cite{bbh1,bbh2}. The gapped surface-states of these insulators were identified as lower-dimensional topological insulators (TI), giving rise to mid-gap states, localized at sharp corners and hinges of a material. The corner-localized states were argued to support topological quantization of multipole moments under suitable boundary conditions.
This fascinating proposal has led to tremendous inter-disciplinary research on HOTIs \cite{langbehn2017,schindler2018b,benalcazar2019,song2017,miert2018,wang2018,kunst2018,dwivedi2018,ezawa2018,you2019,lee2020,vega2019,varjas2019,chen2020,calugaru2019, zhang2021, xie2021}, which can be realized in solid state materials \cite{schindler2018a,jack2019,jack2020}, photonic crystals \cite{noh2018, hassan2019,mittal2019,chen2019, xie2019, kempkes2019, liu2021}, ultracold atomic systems~\cite{bibo2020}, and mechanical meta-materials \cite{imhof2018,peterson2018,garcia2018,ni2019,xue2019a,xue2019b,zhang2019, fan2019, xue2020, ni2020, peterson2020}. 

In spite of exciting ongoing research, the bulk topological invariants of HOTIs are still not clearly understood \cite{bbh1,geier2018,shiozaki2014,trifunovic2019,khalaf2018a,khalaf2018b,schindler2019,ahn2019,okuma2019,roberts2020, roy2021}. Benalcazar \emph{et al.} proposed the following topological properties of Wilson loops (WL) of non-Abelian Berry connections of HOTIs. (i) As functions of transverse momentum components, the gauge-invariant eigenvalues of WLs along high-symmetry axes (or Wannier centers) do not exhibit any non-trivial winding. (ii) The generators of WLs or Wilson loop Hamiltonians (WLH) display gapped spectra, since the mirror symmetry operators along different principal axes do not commute. (iii) The gauge-dependent WLHs of $D$-dimensional HOTIs are examples of $(D-1)$-dimensional TIs, which can be classified within the paradigm of ``nested Wilson loops" (NWL), by computing Wilson loops of Wilson loops. The NWL is a tool for dimensional reduction and topological classification of gauge-fixing ambiguities of non-Abelian Berry connections on a periodic manifold, like Brillouin zone (BZ) torus.  

Since it cannot be easily implemented for complex \emph{ab initio} band-structures of real materials, the solid-state candidates of HOTI are identified by employing (i) complimentary analysis of various symmetry indicators and spectra of WLs under periodic boundary conditions (PBC)~\cite{khalaf2018a,khalaf2018b,bouhon2019}, and (ii) direct calculations of surface- and corner- states under open boundary conditions (OBC)~\cite{langbehn2017}. 
The spectra of WLs for elemental bismuth have been shown to possess strongly direction-sensitive behavior. 
Thus, elemental bismuth has been identified as a \emph{mixed-order topological insulator}, which can exhibit gapless or gapped surface-states, depending on the orientation of surface~\cite{schindler2018a,schindler2018b,hsu2019}. 
The direction-sensitive gapless surface-states also occur for anti-ferromagnetic topological insulators~\cite{mong2010,otrokov2019, liMnBiTe, chenMnBiTe, haoMnBiTe} and many topological crystalline insulators~\cite{po2017, bradlyn2017, chen2017, slager2013, kruthoff2017}. 
However, in real materials, due to the complexity of band structures, the precise relationship between surface-states and corner-states cannot be clearly addressed. 

This has motivated us to ask the following questions for analytically controlled BBH models of $D$-th order TIs. (i) Can the WLs of BBH models support gapless spectra? (ii) Do BBH models support gapless surface-states along any directions? In this work, we perform explicit analytical and numerical calculations to affirmatively answer these questions and establish mixed-order topology of BBH models. We show the WLs along body-diagonal directions of cubic Brillouin zone possess gapless spectra and the gapless surface-states along body diagonals correspond to $2^{D-1}$-component massless, Dirac fermions. For two-dimensional BBH model of quadrupolar TIs, we also show the corner-states arise, when the flow of Dirac fermions is obstructed by boundary conditions.

\begin{figure*}[!]
\centering
\includegraphics[width=0.9\textwidth]{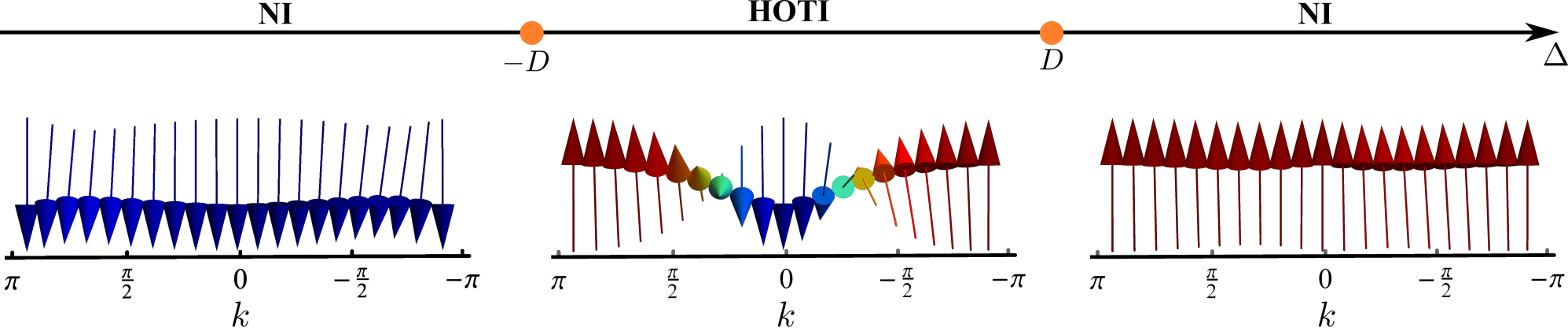}
\caption{The phase diagram of $D$-dimensional, $D$-th order topological insulator, described by BBH model of Eq.~\ref{eq:BBH}. 
Topologically distinct phases are separated by quantum critical points at $|\Delta| = D$ (marked by colored dots), where bulk band gap vanishes at either the zone center or corner, thereby realizing a single, bulk Dirac point. 
The topologically trivial (NI) phase is distinguished from the higher-order topological insulator (HOTI) by the integer winding number of the unit vector $\hat{\bs{n}}$ from Eq.~\ref{Eq:4} along the body diagonal direction of cubic Brillouin zone (shown by the texture of the arrows). 
Thus the difference between the NI and the HOTI can be concisely understood in terms of the first homotopy classification of Bloch Hamiltonian, along the high-symmetry lines, as proposed in the present work.
}
\label{fig:topology}
\end{figure*}

\section*{Models and phase diagrams}
Minimal models of \emph{cubic symmetry preserving} $D$-th order TIs are constructed by combining one extended $s$-wave function $f_s(\bs{k})=[\Delta-\sum_{j=1}^{D} \cos(k_j)]$, all independent $p$-wave harmonics $f^j_p(\bs{k})=\sin(k_j)$, and \emph{a specific class of $d$-wave harmonics}, $f^{l}_d(\bs{k})=\sqrt{\frac{2}{l(l+1)}}\left[ l \cos(k_{l+1}) - \sum_{s=1}^{l} \cos(k_s) \right]$ with $l = 1,2,..(D-1)$, as elements of a $2D$-dimensional vector field,
\begin{align}
\bs{N}(\bs{k})&=[t_p f^1_p(\bs{k}),..,t_p f^D_p(\bs{k}), t_d f^1_d(\bs{k}),.., t_d f^{D-1}_d(\bs{k}), \nonumber \\ 
&  \qquad t_s f_s(\bs{k})], 
\label{eq:BBH-N}
\end{align} 
where $t_s$, $t_p$, and $t_d$ are three independent hopping parameters.
Note that the role of such $d$-wave terms is to reduce the number of TRIM points, supporting band inversion.
The Bloch Hamiltonian 
 \begin{align}
\hat{H}_D = t_p \sum_{j=1}^{D}  f^j_p(\bs{k}) \Gamma_j +  t_d \sum_{j=1}^{D-1} f^j_d(\bs{k}) \Gamma_{D+j} + t_s f_s(\bs{k}) \Gamma_{2D}.  
\label{eq:BBH}
\end{align} 
operates on $2^{D}$-component spinor $\Psi(\bs{k})$, and $\Gamma_j$'s are mutually anti-commuting, $2^D \times 2^D$ matrices. 
Since the absent gamma matrix $\Gamma_{2D+1}$ anti-commutes with $\hat{H}_D(\bs{k})$, the $D$-th order HOTI exhibits $\mathbb{Z}_2$ particle-hole symmetry.
The operator $\mathcal{O}(\bs{k})=\Psi^\dagger(\bs{k}) \Gamma_{2D+1} \Psi(\bs{k})$ has the irreducible representation of $2^D$-th multipole moment of $D$-dimensional cubic point group, which behaves as $\prod_{i=1}^{D} x_i$. 
For $D=2,3$, these correspond to $B_{2}$ quadrupole and $A_{2}$ octupole, respectively.
The specific forms of BBH models~\cite{bbh1,bbh2} can be obtained from $\hat{H}_D(\bs{k})$ with special choices of hopping parameters and $\Delta$, viz. for $D=2$ [$D=3$] the parameters  $(t_s, t_p, t_d, \Delta) = (-\lambda/\sqrt{2},  \lambda, \lambda/\sqrt{2}, -2\gamma/\lambda)$ [$(-\lambda/\sqrt{3}, \lambda, \lambda/\sqrt{2}, 0)$]. 
By setting $t_d=0$, we recover models of first-order TIs, with a continuous $O(D)$ symmetry.


The spectra of $2^{D-1}$-fold degenerate conduction and valence bands are given by $E_{\pm}(\bs{k})=\pm |\bs{N}(\bs{k})|$ and the phase diagram is illustrated in Fig.~\ref{fig:topology}. 
In the parameter regime $|\Delta|<D$, the bands of $D$-th order HOTI are inverted between the center $\bs{k}=(0,0,..,0)$ and the corner $\bs{k}=(\pi,\pi,..,\pi)$ of the cubic BZ with respect to the matrix $\Gamma_{2D}$, giving rise to a topologically non-trivial phase. The presence of band inversion at two TRIM points is due to the presence of $d$-wave harmonic $f^{D-1}_d(\bs{k})$.
The trivial phases (NI) occurring for $|\Delta|>D$ are separated from the non-trivial phase by topological quantum phase transitions at $\Delta=\pm D$. 
At these critical values of $\Delta$, the spectral gap respectively vanishes at $\bs{k}=(0,0,..,0)$ and $\bs{k}=(\pi,\pi,..,\pi)$, while all other TRIM points remain gapped in the entire phase diagram. 
Therefore, the universality class of topological phase transition between a $D$-th order TI and NI is described by one species of $2^D$-component, massless Dirac fermion.
By contrast, lower order TIs can support additional topologically non-trivial states, separated by band-gap closing at other TRIM points.  

\section*{Bulk topological invariants}
While the physical properties of $\hat{H}_D$ are independent of representations of $\Gamma$ matrices, we will assume $\Gamma_{2D}$ to be a diagonal matrix, and other anti-commuting matrices will be chosen to have block off-diagonal forms. The projection operators of conduction and valence bands are determined by $O(2D)$ unit vector $\hat{N}(\bs{k})$, which lies on a unit sphere $S^{n}$, with $n=2D-1$. Since, $\pi_D(S^n)$ is trivial for any $n>D$, the topology of HOTIs cannot be straightforwardly described in terms of spherical homotopy classification of $\hat{N}(\bs{k})$. However, progress can be made by treating $\hat{H}_D(\bs{k})$ as a non-uniform order parameter ($\bs{k}$ dependent texture) that describes a pattern of symmetry breaking $SO(2D) \to SO(2D-1)$. Thus, $\hat{H}_D(\bs{k})$ defines map from the space group of a $D$-cube to the coset space $\frac{SO(2D)}{SO(2D-1)}= \frac{Spin(2D)}{Spin(2D-1)}=S^{2D-1}$. On fermionic spinor $\Psi$, the action of special orthogonal group $SO(m)$ is realized in terms of its universal double cover group $Spin(m)$, such that $SO(m)=Spin(m)/\mathbb{Z}_2$. Consequently, $\hat{H}_D(\bs{k})$ is diagonalized by unitary transformation 
\begin{equation}
U^\dagger(\bs{k}) \hat{H}_D(\bs{k}) U(\bs{k}) = |\bs{N}(\bs{k})| \Gamma_{2D}.
\end{equation}
Since the conduction and valence bands are $2^{D-1}$-fold degenerate, the diagonalizing matrix $U(\bs{k})$ can only be determined up to $Spin(2D-1)$ gauge transformation and $U(\bs{k}) \in \frac{Spin(2D)}{Spin(2D-1)}$. Consequently, the gauge group of intra-band Berry's connection corresponds to $SO(2D-1)=\frac{Spin(2D-1)}{\mathbb{Z}_2}$. A convenient form of intra-band connection is given by
\begin{eqnarray}
A_\mu(\bs{k}) =\frac{(N_a \partial_\mu N_b - N_b \partial_\mu N_a)}{4(|\bs{N}| + N_{2D})} \Gamma_{ab},\label{connection1}
\end{eqnarray}
where $\Gamma_{ab}=\frac{1}{2i} [\Gamma_a, \Gamma_b]$, with $a=1,.., 2D-1$, and $b=1,..,2D-1$ are the generators of $SO(2D-1)$ group. By acting with the projection operators $P_\pm = \frac{1}{2} (1 \pm \Gamma_{2D})$, one arrives at Berry connections for conduction ($+$) and valence ($-$) bands.

Due to the underlying cubic symmetry (symmetry of base manifold), various components of $\bs{N}(\bs{k})$ will vanish at high-symmetry points, and on high-symmetry axes and planes, indicating partial restoration of global symmetries. 
Therefore, high-symmetry points, lines and planes serve as topological defects of the $SO(2D)$-vector field and $SO(2D-1)$ intra-band Berry connection. 
Since at the band-inversion points, $\bs{N}$ has only one non-vanishing component, they support maximal restoration of the $SO(2D-1)$ global symmetry. 
Hence, these TRIM points will be mapped on to the north- and south-poles of $S^{2D-1}$, causing Dirac-string singularities of the Berry connection described by Eq.~\ref{connection1}. 
In contrast to this, on high-symmetry axes and planes smaller sub-groups of $SO(2D)$ are restored.
The corresponding defects may be classified by one and two dimensional winding numbers. Among all high-symmetry lines (planes), the body diagonal axes (dihedral planes) connecting the points of band inversion show the maximal restoration of global symmetry and play essential roles toward topological classification.   

\begin{figure}[!t]
\centering
\subfloat[\label{fig:foti2D}]{%
  \includegraphics[width=0.45\columnwidth]{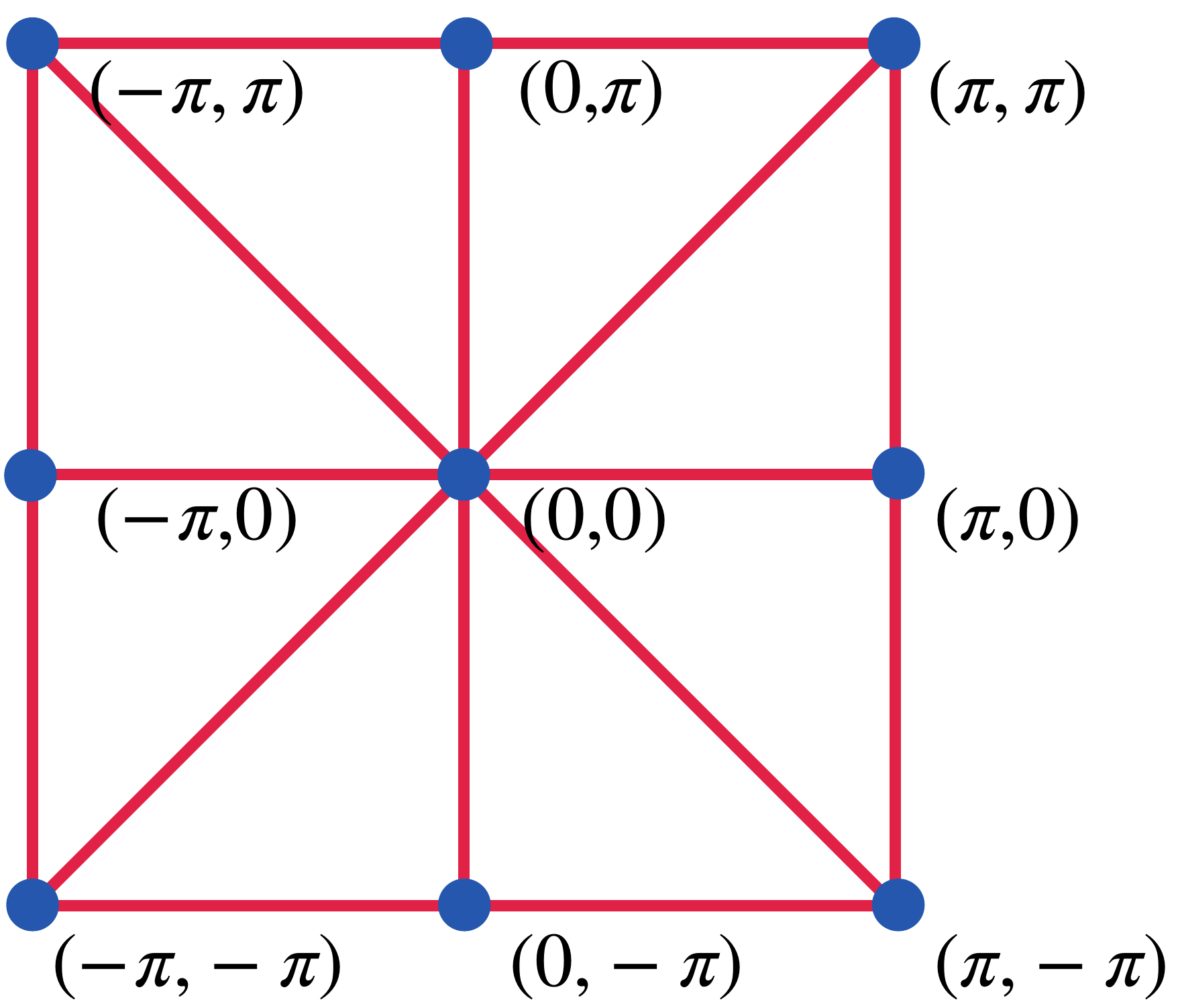}%
}
\hfill
\subfloat[\label{fig:hoti2D}]{%
  \includegraphics[width=0.45\columnwidth]{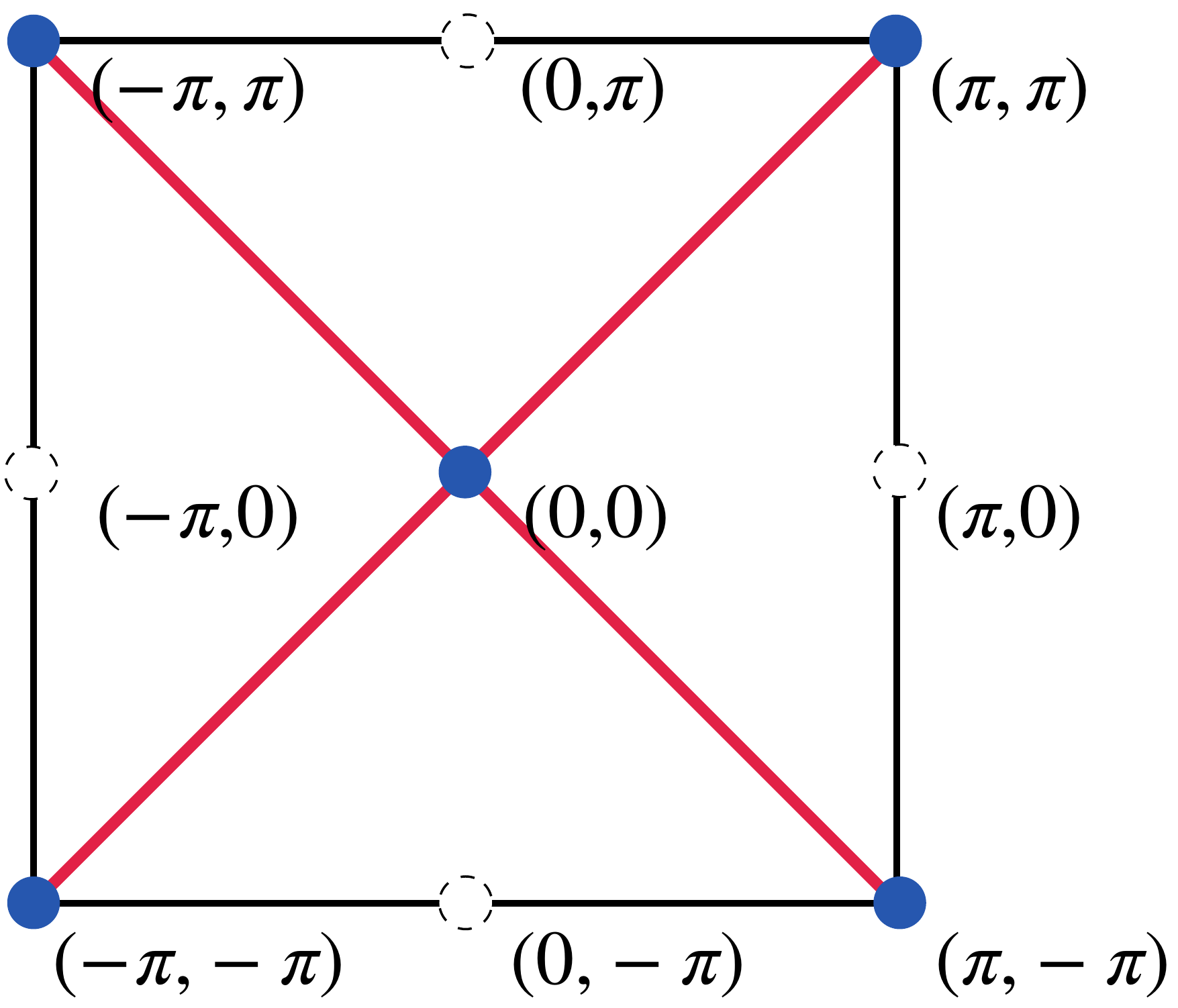}%
}
\hfill
\caption{Distinction between two-dimensional, first- and second- order topological insulators, based on the number of high-symmetry lines in the Brillouin zone (BZ) that can be classified by the fundamental group of a circle, $\pi_1(S^1)$ [red lines]. 
(a) For first-order topological insulators (TIs), the Bloch Hamiltonian maps each line connecting two high-symmetry points (blue dots) in the BZ to a circle in the Hilbert space. 
Thus, all such mirror lines may be topologically classified by $\pi_1(S^1)$.
(b) In contrast to this, for a second-order TI, exemplified by the Benalcazar-Bernevig-Hughes model, only the diagonal mirror lines can be classified by $\pi_1(S^1)$.
}
\label{fig:1st&D}
\end{figure}

\subsection*{One-dimensional winding number}
Irrespective of the spatial dimensionality, all mirror-symmetry preserving $d$-wave harmonics ($f^j_d$) vanish along the body diagonal-axes. 
The Bloch Hamiltonian along these lines is described by two-component vectors, indicating restoration of $SO(2D-2)$ global symmetry. 
For example, by setting $k_x=k_y=...=k_D=k$, we arrive at the following two-component model 
\begin{align}
\hat{H}_D(k)= t_p \; \sqrt{D} \; \sin(k) \Gamma^\prime + [\Delta-D \cos(k)]\Gamma_{2D} \label{Eq:3}
\end{align}
along the $[ 1, 1,..,1]$ direction, with $ \Gamma^\prime =( \sum_{j=1}^{D} \; \Gamma_j)/\sqrt{D}$.
Owing to this, the two-component unit vector 
\begin{align}
\hat{\bs{n}}(k) = \frac{(t_p \sqrt{D} \sin k,  \Delta-D \cos k)}{\sqrt{D t_p^2 \sin^2 k + (\Delta - D \cos k )^2}} \label{Eq:4}
\end{align}
describes map from a non-contractible cycle ($S^1$) of the Brillouin zone to a coset space $S^1$, with a non-trivial winding number.
When $|\Delta|<D$,  $\hat{H}_D(k)$ describes a one-dimensional TI, classified by the fundamental group of a circle $\pi_1(S^1)=\mathbb{Z}$, which is captured by the winding of the angle 
\begin{align}
\theta(k) = \tan^{-1}\frac{t_p \sqrt{D} \sin k}{ \Delta-D \cos k},
\end{align}
as shown in Fig. \ref{fig:topology}. 
Since we are working with toy models with only nearest neighbor hopping, the winding number can only take values $0, \pm 1$. 
By introducing further neighbor hopping terms one can find higher winding numbers~\cite{wang2022}. 
In contrast to HOTIs, first-order TIs support band inversion at all TRIM points, \emph{and} all high-symmetry lines can be classified by $\pi_1(S^1)$.
This distinction between first and $D$-th order TIs at $D=2$ is summarized in Fig.~\ref{fig:1st&D}.

\subsection*{Two-dimensional winding number}
The two-dimensional (three-dimensional) HOTI also supports a rotational symmetry protected, quantized non-Abelian Berry flux through its entire bulk (threefold-rotation  symmetric or $C_3$ and dihedral planes).
Because the two-dimensional HOTI is a special case of the planes in $C_4$ rotational-symmetry protected Dirac semimetals that lie in-between the Dirac points, it can be classified by a pair of two-dimensional winding numbers~\cite{tyner2020}. The quantized flux will be associated with a specific component of the generator of $C_4$ symmetry, which is given by the commutator of mirror operators $\hat{M}_x$ and $\hat{M}_y$, i.e., $[\hat{M}_x,\hat{M}_y]=\Gamma_{12}$.
Here, we do not discuss it further; instead, we focus on the dihedral and $C_3$ planes of three-dimensional 3rd order TIs, which are planes perpendicular to two out of three $n$-fold rotational axes ($n = 2, 3,4$) supported by the $O_h$ point group. 
In particular, we show that these planes carry quantized non-Abelian,  Berry flux, which allows us to identify two-dimensional BBH models, and both $C_3$ and  dihedral planes of three-dimensional BBH models as generalized, quantum spin Hall insulators~\cite{tyner2020}.
The remaining set of planes perpendicular to the principle axes possess fourfold-rotational symmetry, and do not support quantized flux.

\begin{figure}[!t]
\centering
\subfloat[\label{fig:dihedral}]{%
  \includegraphics[width=0.47\columnwidth]{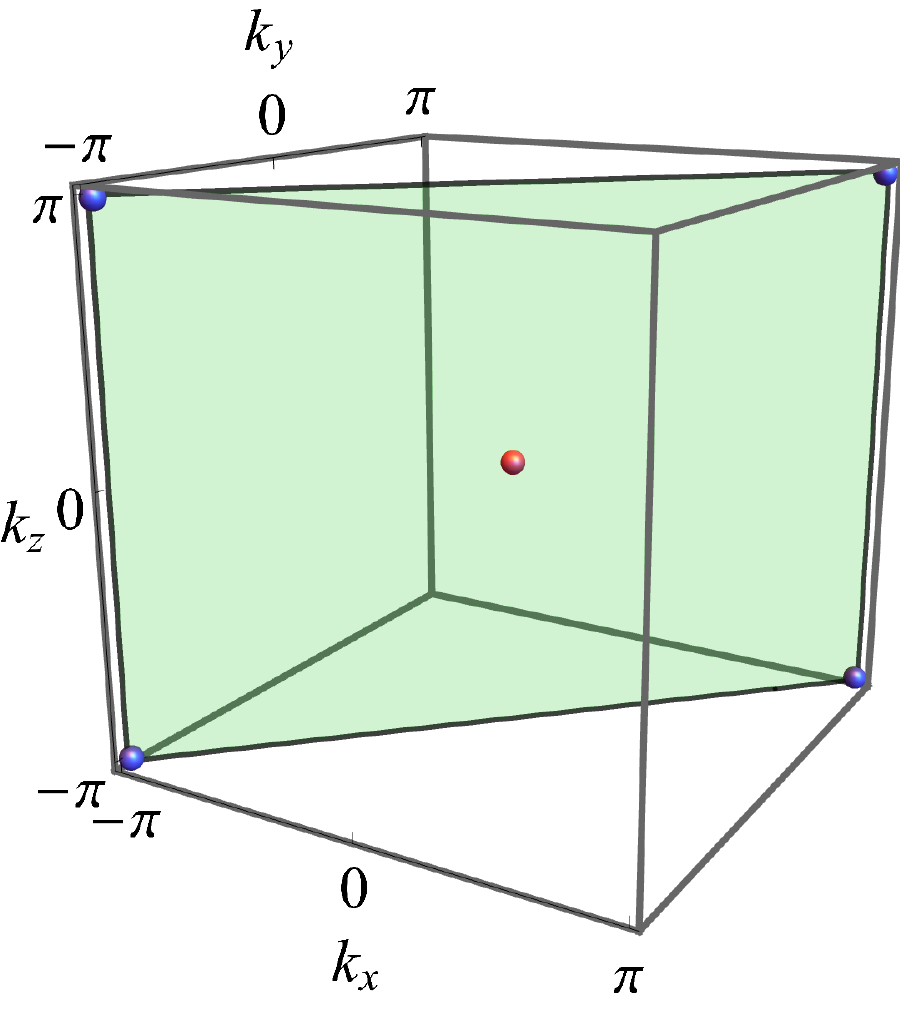}%
}
\hfill
\subfloat[\label{fig:111}]{%
  \includegraphics[width=0.47\columnwidth]{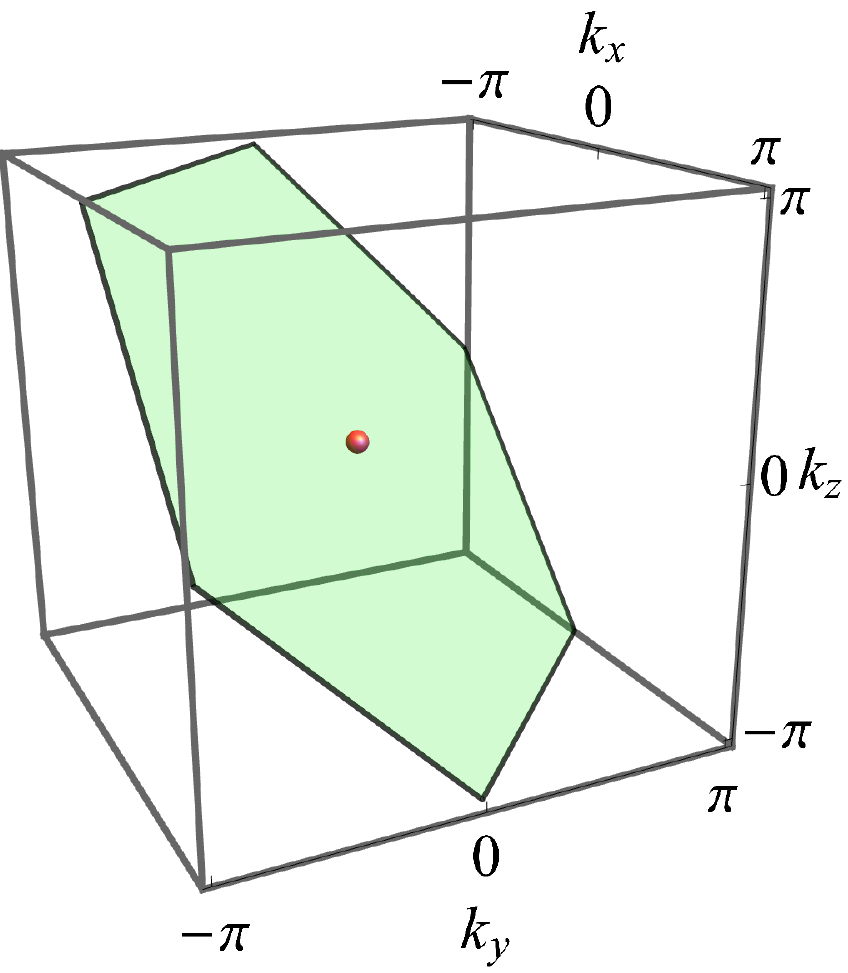}%
}
\hfill
\caption{Planes in the three-dimensional cubic Brillouin zone supporting quantized non-Abelian Berry flux of magnitude $2 \pi$. 
(a) One of the six dihedral planes (shaded) with the red and blue dots representing high-symmetry points that support band-inversion. 
The Bloch Hamiltonian projected to  this plane has the form of two-dimensional, $C_2$-symmetric, Benalcazar-Bernevig-Hughes (BBH) model [see \eqref{dihedralplane} ]. 
(b) By contrast, planes perpendicular to the body diagonals are $C_3$ symmetric, and support a single band-inversion at the zone center. 
The corresponding projected Bloch Hamiltonian, given by \eqref{eq:H-111}, is unrelated to the BBH model. 
}
\label{fig:flux-planes}
\end{figure}

If we consider the high-symmetry plane, which is perpendicular to $[1, -1, 0]$ axis (exemplified by the $k_x = k_y$ plane in Fig.~\ref{fig:dihedral} ), the Bloch Hamiltonian of 3rd-order TI will be reduced to two-dimensional, $C_2$-symmetric HOTl, written with $8 \times 8$ anti-commuting matrices. 
In the rotated basis, $k_\pm=(k_x \pm k_y)/\sqrt{2}$, the Hamiltonian for $[1, -1,0]$ plane will become 
\begin{align}
&\hat{H}_3(k_+, k_-=0,k_z) = \sqrt{2} t_p \sin (\frac{k_+}{\sqrt{2}}) \Gamma_+ + t_p \sin (k_z) \Gamma_3 \nn \\
&+ \frac{2t_d}{\sqrt{3}} \left[\cos(k_z) -\cos(\frac{k_+}{\sqrt{2}}) \right]\Gamma_5  
+ t_s \Bigl[\Delta- \cos(k_z) \nn \\ & - 2\cos(\frac{k_+}{\sqrt{2}}) \Bigr]\Gamma_6, \label{dihedralplane}
\end{align}
where $\Gamma_\pm= (\Gamma_1 \pm \Gamma_2)/\sqrt{2}$. 
The Hamiltonian $\hat{H}_3$ is invariant under twofold rotations generated by  $\Gamma_{3+} = \frac{1}{2i}[\Gamma_3, \Gamma_{+}]$.
It can be shown that the projection of the non-Abelian Berry connection on the rotation generator $\Gamma_{3+}$, $\bs A_{3+} \coloneqq \frac{1}{8}\tr{\bs A ~ \Gamma_{3+}}$, supports a Berry flux of magnitude $2\pi$~\cite{tyner2020}. 
We note that all 12 dihedral planes (corresponding to $k_- = 0, \sqrt{2}\pi$ planes)  
carry $2\pi$ fluxes carried by respective twofold-rotation generators.

\begin{figure}[!t]
\centering
\includegraphics[width=0.95\columnwidth]{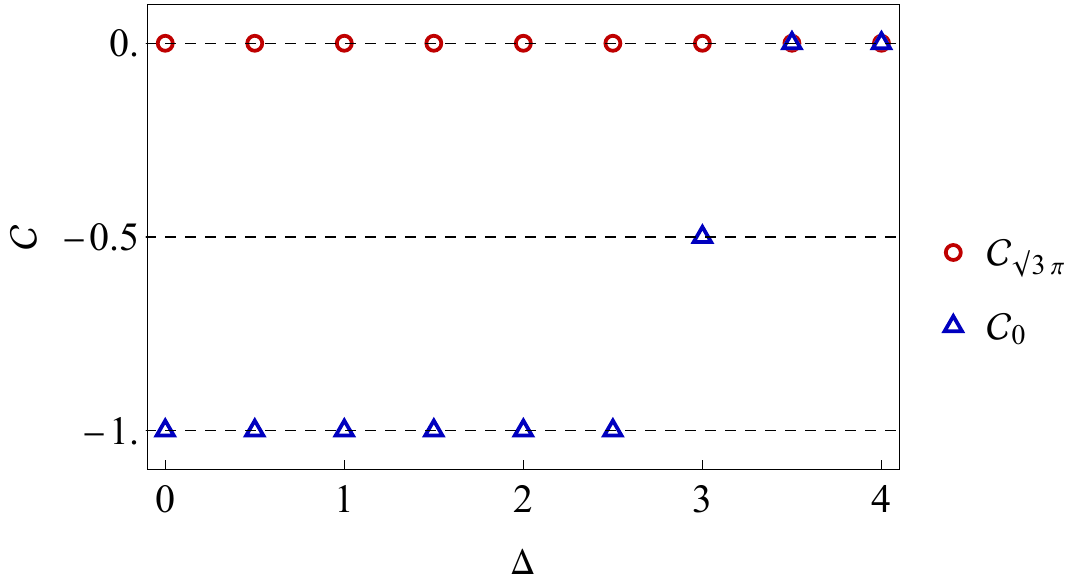}%
\caption{ 
Comparison between the relative Chern numbers carried by the $k_3 = 0$ and $\sqrt{3}\pi$ planes. 
The difference is equivalent to a 3D winding number.
Here, only the higher order TI phase ($|\Delta| < 3$)  supports a finite 3D winding number.
}
\label{fig:C3-Chern}
\end{figure}

In crystals with the cubic symmetry, planes in the BZ  perpendicular to the body-diagonals possess a threefold rotational symmetry.
Such planes support rotational-symmetry protected, non-Abelian, quantized flux in 3rd order TIs.
In order to demonstrate the presence of the quantized flux, we consider the plane perpendicular to the $[1,1,1]$ axis, which takes the form of a hexagonal Brillouin zone, as shown in Fig.~\ref{fig:111}. 
We introduce a set of rotated coordinates,
$(k_1, k_2, k_3) \equiv  \qty(\frac{k_x - k_y}{\sqrt{2}}, \frac{k_x + k_y - 2k_z}{\sqrt{6}}, \frac{k_x + k_y + k_z}{\sqrt{3}} )$, 
to express the reduced Hamiltonian on the $k_3 = 0$ plane as
\begin{align}
& H_{k_3 = 0}(k_1, k_2) = t_p \sin \left(\frac{k_1}{\sqrt{2}}+\frac{k_2}{\sqrt{6}}\right) \Gamma_1 
-\sin{\sqrt{\frac{2}{3}} k_2} ~\Gamma_3 \nn \\
&  - t_p \sin \left(\frac{k_1}{\sqrt{2}}-\frac{k_2}{\sqrt{6}}\right) \Gamma_2 
+ 2 t_d \sin{\frac{k_1}{\sqrt{2}}}  \sin{\frac{k_2}{\sqrt{6}}}~\Gamma_4 \nn \\
& + \frac{2 t_d}{\sqrt{3}} \left[\cos{\sqrt{\frac{2}{3}} k_2} - \cos{\frac{k_1}{\sqrt{2}}} 
\cos{\frac{k_2}{\sqrt{6}}}
\right] \Gamma_5 \nn \\
& + t_s \left[\Delta -2 \cos {\frac{k_1}{\sqrt{2}}} \cos{\frac{k_2}{\sqrt{6}}} - \cos{\sqrt{\frac{2}{3}} k_2}
\right] \Gam_6.
\label{eq:H-111}
\end{align}
Threefold rotations on the $k_3=0$ plane are generated by $\tilde \Gam = (\Gam_{12} + \Gam_{23} + \Gam_{31})/\sqrt{3}$ and $\Gam_{45}$, with the rotation operator $\hat C_3^{[111]} \coloneqq \exp{i \frac{\pi}{3}(\tilde \Gam + \Gam_{45})}$.
In order to obtain the Abelian projected Berry flux threading the $k_3 = 0$ plane, we follow the  methods in Ref.~\cite{tyner2020}, and integrate the Berry curvature supported by the .
First, we deduce the projections of the non-Abelian Berry connection, $\bs A$, on the rotation generators, $\tilde{\bs A} \coloneqq \frac{1}{8}\tr{\bs A~ \tilde \Gamma} $ and $\bs A_{45} \coloneqq \frac{1}{8} \tr{\bs A~  \Gamma_{45}} $.
Next, we integrate $\curl \tilde{\bs A}$ and $\curl \bs A_{45}$ over the $k_3 = 0$ plane to obtain the respective Abelian fluxes. 
We find that $\tilde{\bs A}$ ($\bs A_{45}$) supports a net $-2\pi$ ($0$) flux.
We note that there exists a second  plane perpendicular to the $[1, 1, 1]$ axis at $k_3 = \sqrt{3} \pi$, which shares the same rotation generators.
However, this plane does not support any finite net flux, as shown in Fig.~\ref{fig:C3-Chern}.
By virtue of the cubic symmetry, all such pairs of $C_3$ planes carry the same pattern of quantized flux.

\subsection*{Three-dimensional invariant}
The pattern of quantized flux on the two planes perpendicular to each body diagonal axis reveals a flux-tunneling configuration, whereby the quantized  non-Abelian Berry flux tunnels from $- 2\pi$ (eg. on the $k_3 = 0$ plane) to $0$ (eg. on the $k_3 = \sqrt{3} \pi$ plane) along the body diagonal direction (eg. $[1,1,1]$).
In analogy to the equivalence between the three dimensional winding number and tunneling of mirror Chern number between mirror planes in 1st order TIs~\cite{tyner2021}, here, the  tunneling of non-Abelian flux along the body diagonals implies the existence of a non-trivial three-dimensional winding number  characterizing the bulk topology of the 3rd order TI.
The 3D winding number is defined as $\mc N_{3D} = \mc C_{\sqrt{3} \pi} - \mc C_{0}$, where $\mc C_{a}$ is the net  Berry flux carried by the $a$-th plane in units of $2\pi$~\cite{tyner2021}.
Here, $\mc N_{3D} = 1$, as demonstrated by  Fig.~\ref{fig:C3-Chern}. 
In contrast to 1st order TIs, which support a tunneling configuration along both the principle axes and body-diagonals, in 3rd order TIs the quantized non-Abelian flux tunnels only along body-diagonals.

Through the analyses of 1, 2, and 3 dimensional invariants, we have demonstrated that $D$-th order TIs in $D$ space dimensions share a subset  of band-topological characters of  1st order TIs.
Therefore, they carry a mixed topology, which allows $D$-th order TIs  to behave like 1st order TIs under suitable conditions.
In order to demonstrate these connections explicitly, in the following sections, we will identify the gapless spectra of WLs along the body-diagonal directions and their physical consequence.

\section*{Bulk-boundary correspondence}
The Wilson loop $W_\parallel(\bs{k}_\perp)$ along the $\hat{k}_\parallel$ direction is defined as 
\begin{align}
 W_\parallel(\bs{k}_\perp) = \frac{1}{\mathcal{N}}\; \mathcal{P} \exp [ i \int^{\pi}_{-\pi} A_\parallel(k_\parallel, \bs{k}_\perp) \; dk_\parallel ] 
\label{eq:WL}
 \end{align}
where $\mathcal{P}$ denotes path-ordering and the $(D-1)$-dimensional wave vector $\bs{k}_\perp$ is orthogonal to $\hat{k}_\parallel$, and $\mathcal{N}$ is the rank of gauge group. 
By construction $W_\parallel$ is an element of the gauge group for Berry connections, transforming covariantly under $\bs{k}_\perp$-dependent gauge transformations. 
Since $\hat{\bs{n}}(k)$ winds an integer number of times around $S^1$, the WLs along body-diagonal directions will be mapped to non-trivial $Z_2$ center elements $\pm \mathbbm{1}$ of $Spin(2D-1)$ group. The odd (even and zero) integer winding leads to $-\mathbbm{1}$ element ($\mathbbm{1}$ element), which corresponds to $\pi \mod 2m \pi$ ($0 \mod 2m\pi$) Berry phase.
The non-trivial $\pi$ Berry phase along any high-symmetry axis is known to cause band-touching of WL bands at the projection of the axis on the transverse $(D-1)$-dimensional BZ.
Therefore, for $D$-th order TIs, whether the WLH is gapless or gapped crucially depends on the direction of WL.
In this section, through explicit examples, we connect the direction sensitivity of WLH to similar behavior of surface states.

\subsection*{Quadrupolar model at D=2}
In order to understand the analytical structure of WLs for two-dimensional, quadrupolar insulators $D=2$, let us consider the components of $Spin(3) \equiv SU(2)$ Berry connections along $[1,\pm1]$ directions. Without any loss of generality we can choose our $4 \times 4$ gamma matrices to be $\Gamma_4=\tau_3 \otimes \sigma_0$, $\Gamma_5=\tau_2 \otimes \sigma_0$, $\Gamma_j=\tau_1 \otimes \sigma_j$, with $j=1,2,3$. After using rotated variables $k_\pm \equiv (k_x \pm k_y)/\sqrt{2}$ and the rotated components of Berry connections $A_\pm=(A_x \pm A_y)/\sqrt{2}$ acquire the following form
\begin{align}
A_\pm(\mathbf k) =  \frac{\sqrt{2} t_p}{2N(\mathbf k)[N(\mathbf k)+N_4(\mathbf k)]} \Omega_{\pm}(k_\mp),
\label{eq:aPM}
\end{align}
where $N(\bs k) \equiv |\bs N(\bs k)|$, with $\bs N(\bs k)$ being defined by \eq{eq:BBH-N}, and the gauge dependent matrix $\Omega_{-s}(k_{s})$ is given by
\begin{eqnarray}
\Omega_{-s}(k_{s}) 
&=&s \sin{\frac{k_{s}}{\sqrt{2}}}  \bigg[
t_{d} \sin{\frac{k_{s}}{\sqrt{2}}} ~(\Gamma_{23} - s \Gamma_{31})
\nn \\ &&- t_p \cos{\frac{k_{s}}{\sqrt{2}}} \Gamma_{12}
\bigg].
\end{eqnarray}
with $s=\pm 1$, and $\Gamma_{12}=\tau_0 \otimes \sigma_3$, $\Gamma_{23}=\tau_0 \otimes \sigma_1$, $\Gamma_{31}=\tau_0 \otimes \sigma_2$ are three generators of $SU(2)$ group. With our gauge choice, the conduction and valence bands support identical form of $SU(2)$ Berry connection. 
Since $\Omega_+(k_-)$ is independent of the variable of integration $k_+$, $W_+(k_-)$ can be calculated by performing ordinary one-dimensional integration, without bothering about discrete path-ordering procedure.
Therefore, the WLHs for $[1,\pm1]$ directions are proportional to $\Om_\pm(k_\mp)$ and the eigenstates of $W_\pm(k_\mp)$ correspond to those of $\Om_\pm(k_\mp)$. The eigenvalues of $W_+(k_-)$ are given by $e^{\pm i \lambda_+(k_-)}$, with $\lambda_+(k_-)$ being determined by
\begin{eqnarray}
\lambda_+(k_-)=\tilde{\lambda}_+(k_-)  \int_{-\sqrt{2} \pi}^{\sqrt{2} \pi} dk_+ \; \frac{\sqrt{2} t_p}{2N(\mathbf k)[N(\mathbf k)+N_4(\mathbf k)]}, \nn \\ \label{WLHspectra}
\end{eqnarray}
where $\pm \tilde{\lambda}_+(k_-)$ are eigenvalues of $\Omega_+(k_-)$, with $\tilde{\lambda}_+(k_-)=\sqrt{2t^2_d \sin^4\left(\frac{k_-}{\sqrt{2}}\right) + \frac{t^2_p}{4} \sin^2(\sqrt{2} k_-)}$.

\begin{figure}[!t]
\centering
\subfloat[\label{fig:foti}]{%
  \includegraphics[width=0.75\columnwidth]{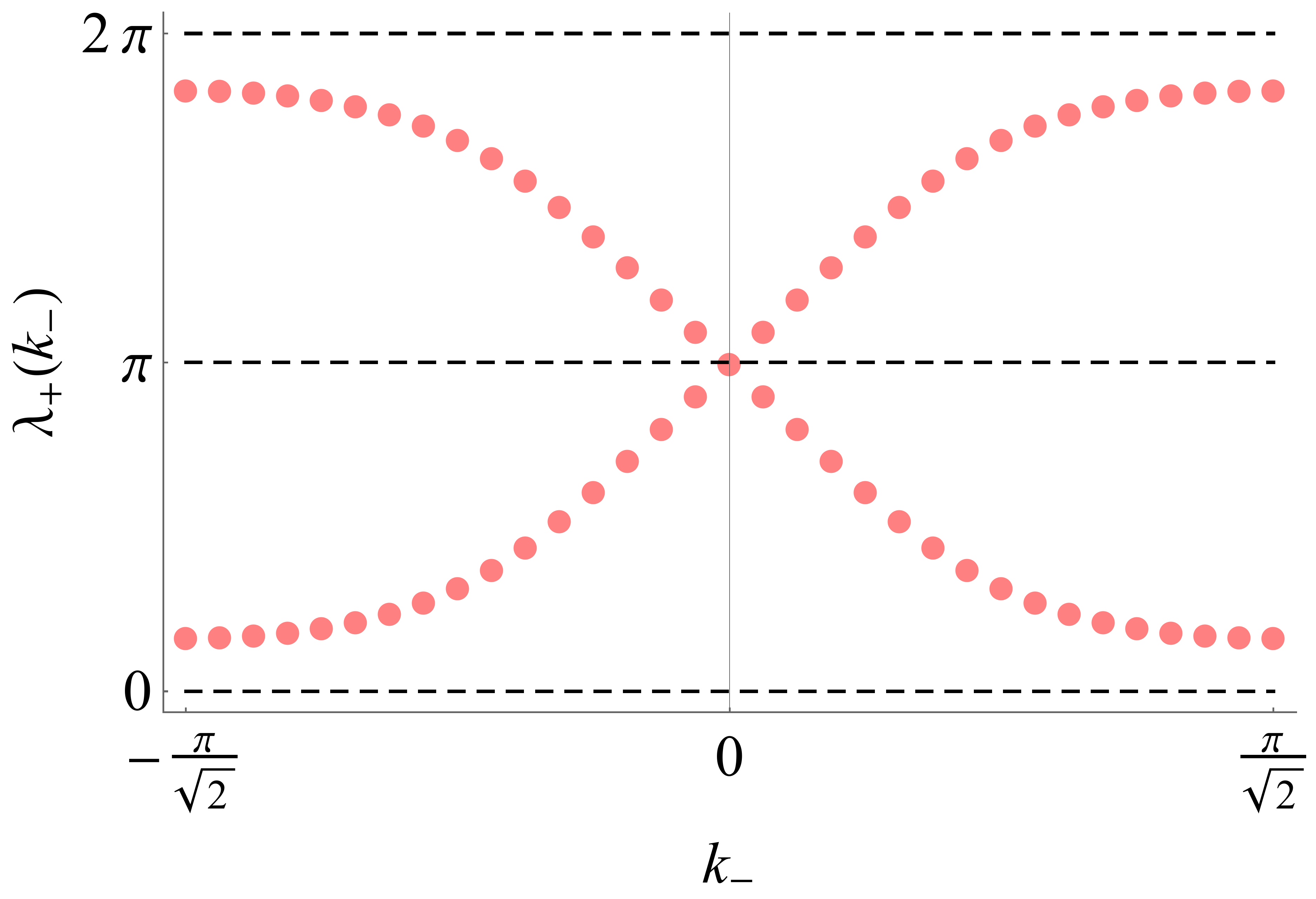}%
}
\hfill
\subfloat[\label{fig:hoti}]{%
  \includegraphics[width=0.75\columnwidth]{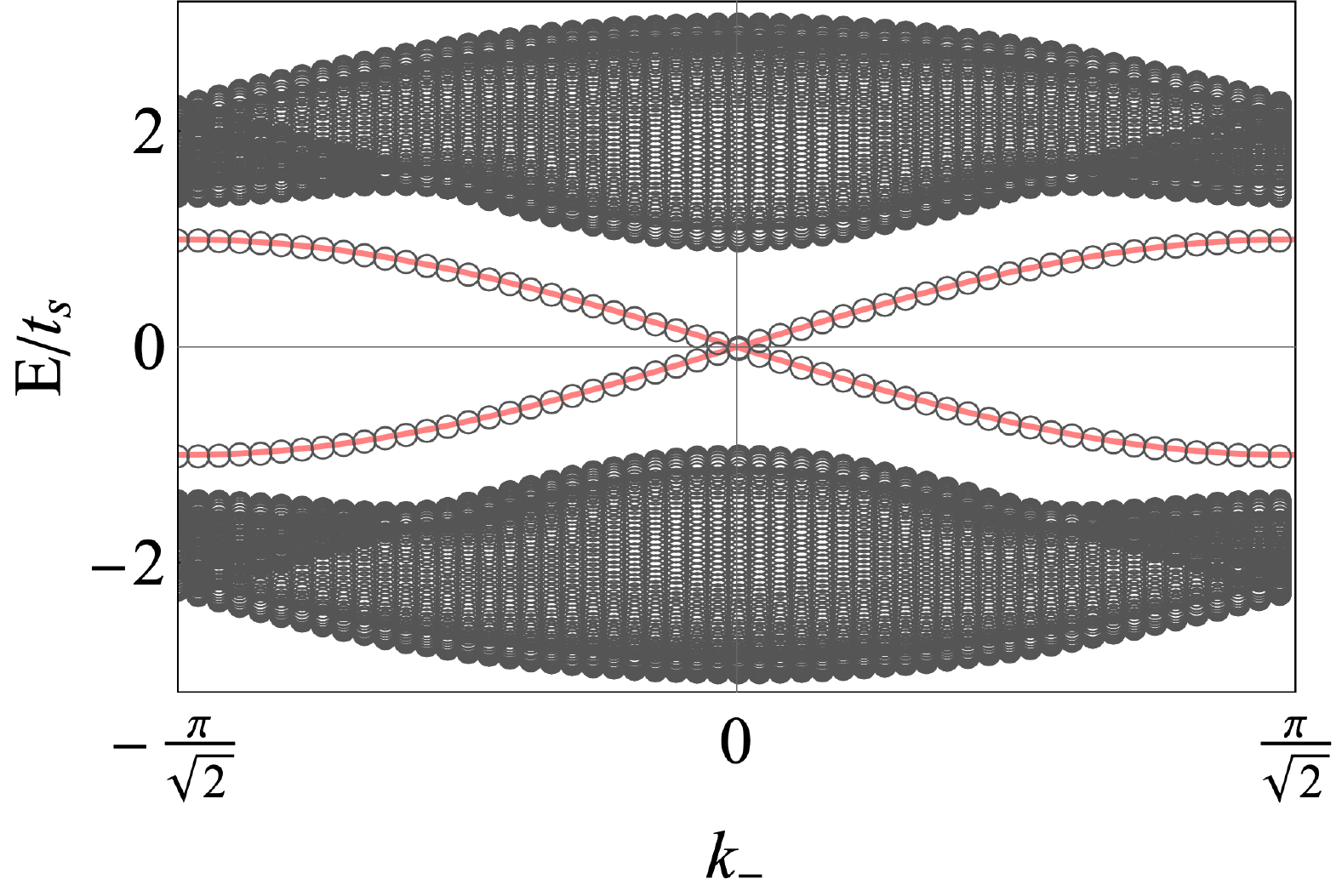}%
}
\hfill
\caption{Gapless spectra of Wilson loops and helical edge-states of two-dimensional, second order topological insulator along $[1,1]$ direction. (a) The gauge invariant spectra of Wilson loop Hamiltonian [Eq.~\ref{WLHspectra} ] support band touching at the center of transverse/surface Brillouin zone.
(b) The spectra of edge-states in the slab geometry with surfaces along the [1,1] direction. 
The (gray) circles represent energy levels obtained by exact diagonalization, while the (red) solid curves are the analytically obtained dispersion [see \eq{eq:diamond-disp}].  
}
\label{fig:(11)}
\end{figure}

For first-order TIs with $t_d = 0$,  $\Om_{+}(k_-)$ vanishes at both $k_- = 0$ and $\pi/\sqrt{2}$. At these singular locations, $W_+(k_-)$ maps to $- \mathbbm{1}$ and $\mathbbm{1}$, respectively. Consequently, the WL spectra become gapless (gapped) at $k_-=0$ ($\pi/\sqrt{2}$), and $\lambda_+(k_-)$ interpolates from $0$ to $2\pi$. In contrast to this, $\Om_{+}(k_-)$ of quadrupolar TIs vanishes only at $k_- = 0$. The $d$-wave term gives rise to non-degenerate eigenvalues of $\Omega_+(k_-)$ at $k_-=\pi/\sqrt{2}$, and $W_+$ never reaches the trivial center element $\mathbbm{1}$.
The gapless behavior of WLH is elucidated in Fig.~\ref{fig:foti}. The isolated singularity of WLH is simultaneously protected by $C_{4v}$ and particle-hole $\mathbb{Z}_2$ symmetries.

\begin{figure}[!t]
\centering
\subfloat[\label{fig:octa}]{%
  \includegraphics[width=0.37\columnwidth]{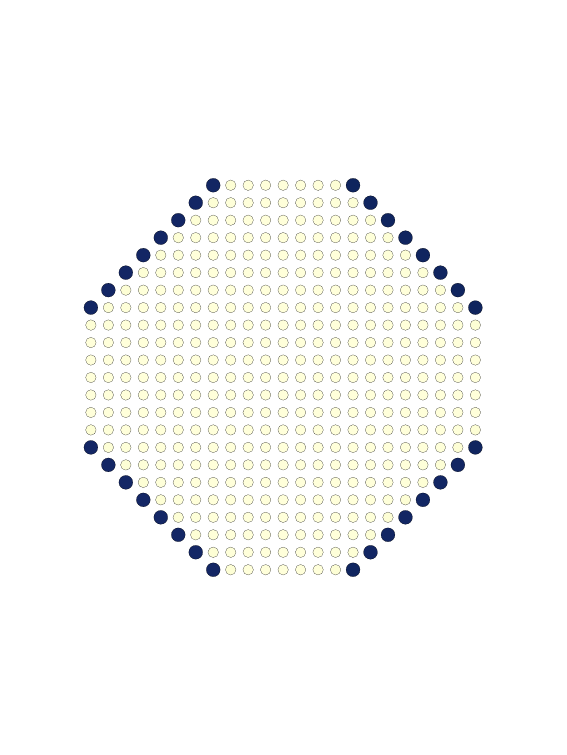}%
}
\hfill
\subfloat[\label{fig:states}]{%
  \includegraphics[width=0.6\columnwidth]{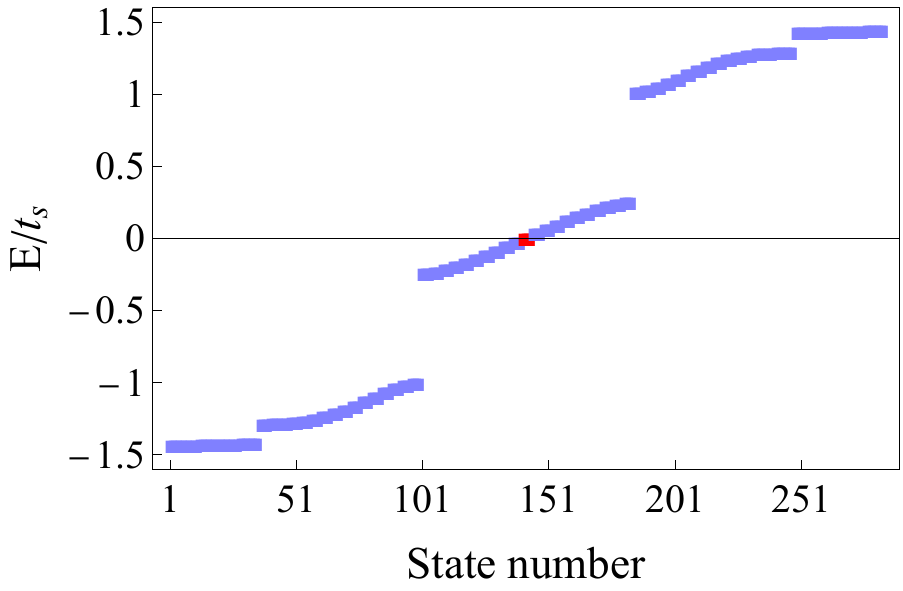}%
}
\hfill
\subfloat[\label{fig:dia}]{%
  \includegraphics[width=0.47\columnwidth]{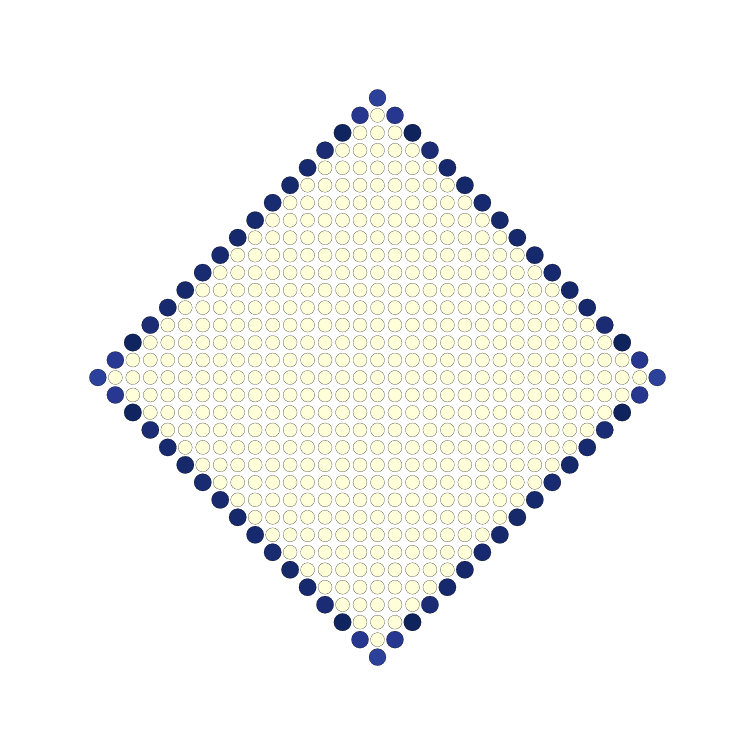}%
}
\hfill
\subfloat[\label{fig:sq}]{%
  \includegraphics[width=0.4\columnwidth]{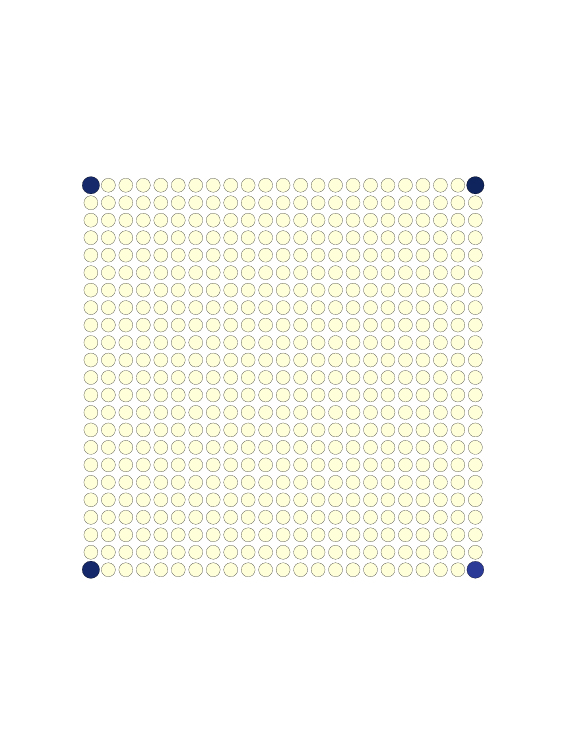}%
}
\hfill
\caption{
Localization pattern of zero-energy states under various geometries. Here, we consider the fourfold symmetric, two-dimensional, 2nd order topological insulator described by  \eq{eq:BBH-N}. 
(a) A flat surface-termination is present perpendicular to all high symmetry axes in the octagonal geometry. Since only the diagonal axes support a non-trivial $\pi_1(S^1)$ winding number, the mid-gap states are solely  localized on the diagonal edges of the octagon.
(b) Energy distribution of the lowest 284 states obtained by exact diagonalization in the octagonal geometry. 
The localization-pattern of the zero energy states (red) is given in (a).
(c) If the diagonal-edges are replaced by corners, such as in the square geometry, corner-localized states are observed.
(d) By contrast, if the $x$ and $y$ edges of the octagon are reduced to points, such as in the diamond geometry, edge propagating modes are obtained. 
}
\label{fig:octagon}
\end{figure}

The manner in which gapless WLH gives rise to gapless edge modes can also be established analytically. 
When the system occupies the half-space $r_+ > 0$, where $r_+$ is the position space conjugate of $k_+$,
the analytical expressions for helical edge modes along $[1, \pm 1]$ directions can be obtained by following Creutz and Horv\'{a}th \cite{creutz1994}. The dispersion for $[1,1]$ edge is given by 
\begin{align}
&\mc E_{[11]}^\pm(k_-) = \pm \frac{ \Delta t_s t_p \sin{\frac{k_-}{\sqrt{2}}}}{\sqrt{t_p^2 \sin^2{\frac{k_-}{\sqrt{2}}} + 2 t_s^2 \cos^2{\frac{k_-}{\sqrt{2}}} } } \nn \\
& ~ \times \Theta\qty(t_p^2 \sin^2{\frac{k_-}{\sqrt{2}}} +2t_s^2 \cos^2{\frac{k_-}{\sqrt{2}}} -  t_s^2 \left| \Delta  \cos{\frac{k_-}{\sqrt{2}}} \right|),
\label{eq:diamond-disp}
\end{align}
where the Heaviside theta function, $\Theta(x)$, implements the condition for normalizability of the surface-states.
For $|k_-| \ll 1$ the dispersion is linear, and describes two counter propagating modes. 
In Fig.~\ref{fig:hoti}, we corroborate the analytically obtained egde-states dispersion by an exact diagonalization of Hamiltonian on a finite cylinder.

Our detailed analysis of two-dimensional HOTI clearly reveals the following mixed-order topological properties: the bulk is a generalized quantum spin Hall state that supports (i) quantized, non-Abelian Berry flux of magnitude $2\pi$; (ii) the presence of gapless (gapped) WLHs along $[1, \pm 1]$ ($[1,0]$ and $[0,1]$) directions; and (iii) the existence of gapless  helical (gapped) edge-modes along $[1, \pm 1]$ ($[1,0]$ and $[0,1]$) directions. Next we demonstrate that the corner-states emerge, when the flow of helical edge-states is  hindered by boundary conditions.

In finite samples, the localization pattern of the topologically protected mid-gap states is strongly geometry dependent. 
For any $D$-dimensional point group,  there exists a sample-geometry with  $(D-1)$-dimensional surface terminations perpendicular to every   high-symmetry axes.
The localization pattern of the mid-gap states in such a geometry  bears a direct correspondence with the $\pi_1(S^1)$-based classification of the bulk topology.
In particular, \emph{if a high-symmetry axis supports a non-trivial $\pi_1(S^1)$ winding number under periodic boundary condition, then the surface-terminations perpendicular to it will support $(D-1)$-dimensional edge-localized states}.
In Fig.~\ref{fig:octagon}(a), we demonstrate this principle through  the localization pattern of the mid-gap states in the two-dimensional 2nd-order TI in an octagonal geometry.
Because the diagonal-axes carry   non-trivial $\pi_1(S^1)$ windings, the zero-energy states [see Fig.~\ref{fig:octagon}(b)] are localized  only on the diagonal edges of the octagon.
If the $x$ and $y$ [diagonal] edges are reduced to points, then the  zero-energy states are localized along the edges [at the corners] of the resultant diamond- [square-] shaped samples, as shown in Fig.~\ref{fig:octagon}(c)[(d)]. 
Therefore, in the diamond geometry, the two-dimensional 2nd-order TI supports propagating edge-localized states. 
It is, thus, notable that the mid-gap states in an HOTI may behave like those in 1st-order TIs under suitable sample geometries. In the following section, we identify the analytical structure of WLH and gapless surface-states at $D=3$.

\subsection*{Octupolar model at D=3}

The $\pi_1(S^1)$ classification of $[1,1,1]$ axis [see Eq. ~\ref{Eq:4} ] and $C_2$-symmetric BBH form at $[1,-1,0]$ dihedral plane [see Eq.~\ref{dihedralplane} ] indicate the presence of gapless WL bands and surface-states. Without any loss of generality, we will use the following representation of $8 \times 8$ gamma matrices: $\Gamma_6=\eta_3 \mathbbm{1}_{4\times 4}$, $\Gamma_7=\eta_2 \otimes \mathbbm{1}_{4 \times 4}$, and $\Gamma_j= \eta_1 \otimes \gamma_j$, where $\gamma_j$'s are five mutually anticommuting $4 \times 4$ matrices. For calculating WL of $SO(5)$ Berry connection along $[1,1,1]$ direction, it is convenient to use the rotated coordinates $(k_1, k_2, k_3)$,  and the rotated component $A_3=(A_x+A_y+A_z)/\sqrt{3}$ of Berry connection, whose form is given by 
\begin{align}
& A_3(\bs k)  = \frac{1}{2N(\bs k)[N(\bs k) + N_6(\bs k)]} \Omega_3(k_1, k_2).
\label{eq:a3}
\end{align}
The WLH will be proportional to the gauge dependent matrix
\begin{align}
& \Omega_3(k_1, k_2)= \frac{1}{\sqrt{3}} 
\sum_{a < b = 2}^5 \omega_{ab}(k_1, k_2) \Gamma_{ab},\label{SO5}
\end{align}
which is independent of $k_3$. 
The explicit expressions of  $\omega_{ab}(k_1, k_2)$'s are presented in Appendix~\ref{app:Om}.
Since $\Om_3$ is independent of $k_3$, the WL is easily obtained by performing one-dimensional integration over $k_3$, and the WLH is proportional to $\Om_3$. The $8\times 8$ matrices $\Gam_{ab} = [\Gam_a, \Gam_b]/(2i)$ correspond to ten generators of $SO(5)$ group, and they can also be expressed as $\Gam_{ab}=\eta_0 \otimes \gamma_{ab}$, where $\gamma_{ab}$ are $4 \times 4$ generators of $SO(5)$ group.
Therefore, identical form of Berry connection for conduction and valence bands will be found by acting with projectors $P_\pm=\frac{1}{2} (1 \pm \Gamma_6)$. The projections of $\Omega_{3}(k_1,k_2)$ on conduction and valence bands can be obtained from Eq.~\ref{SO5}, with the replacement $\Gamma_{ab} \to \gamma_{ab}$. Although all $SO(5)$ generators simultaneously appear in $\Omega_{3}(k_1,k_2)$, it can be shown that $\Omega^2_{3}(k_1,k_2) \propto \mathbbm{1}$. Consequently, at any generic location of $(k_1,k_2)$-plane, the eigenvalues of WLH will be two-fold degenerate. Since the eigenstates of $\Omega_3(k_1,k_2)$ are also the eigenstates of $W_3(k_1,k_2)$, $\Omega_3(k_1,k_2)$ describes $SO(5)/SO(4)$ gauge fixing of Berry connection.

\begin{figure}[!t]
\centering
\subfloat[\label{fig:foti1}]{%
  \includegraphics[width=0.75\columnwidth]{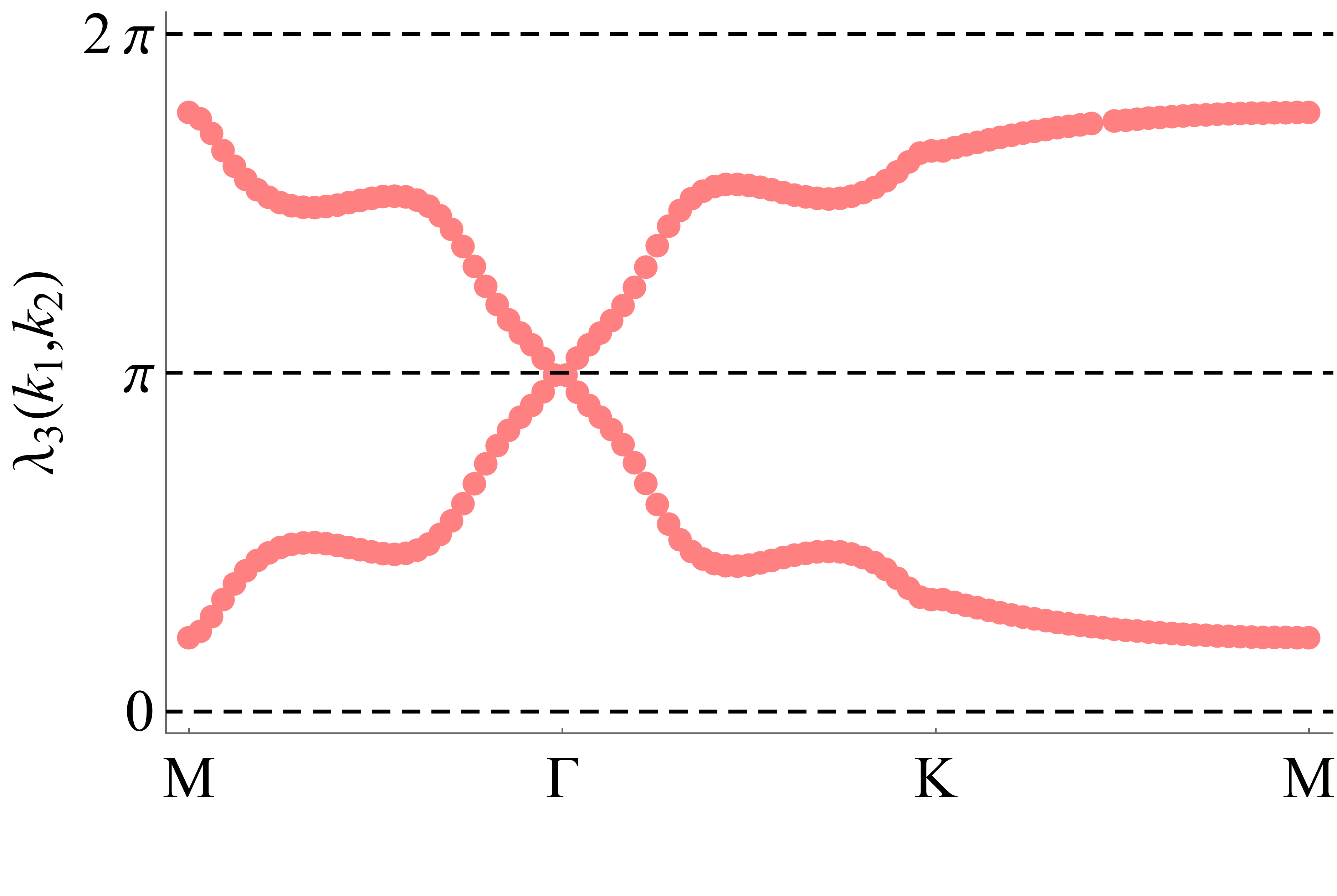}%
}
\hfill
\subfloat[\label{fig:hoti1}]{%
  \includegraphics[width=0.75\columnwidth]{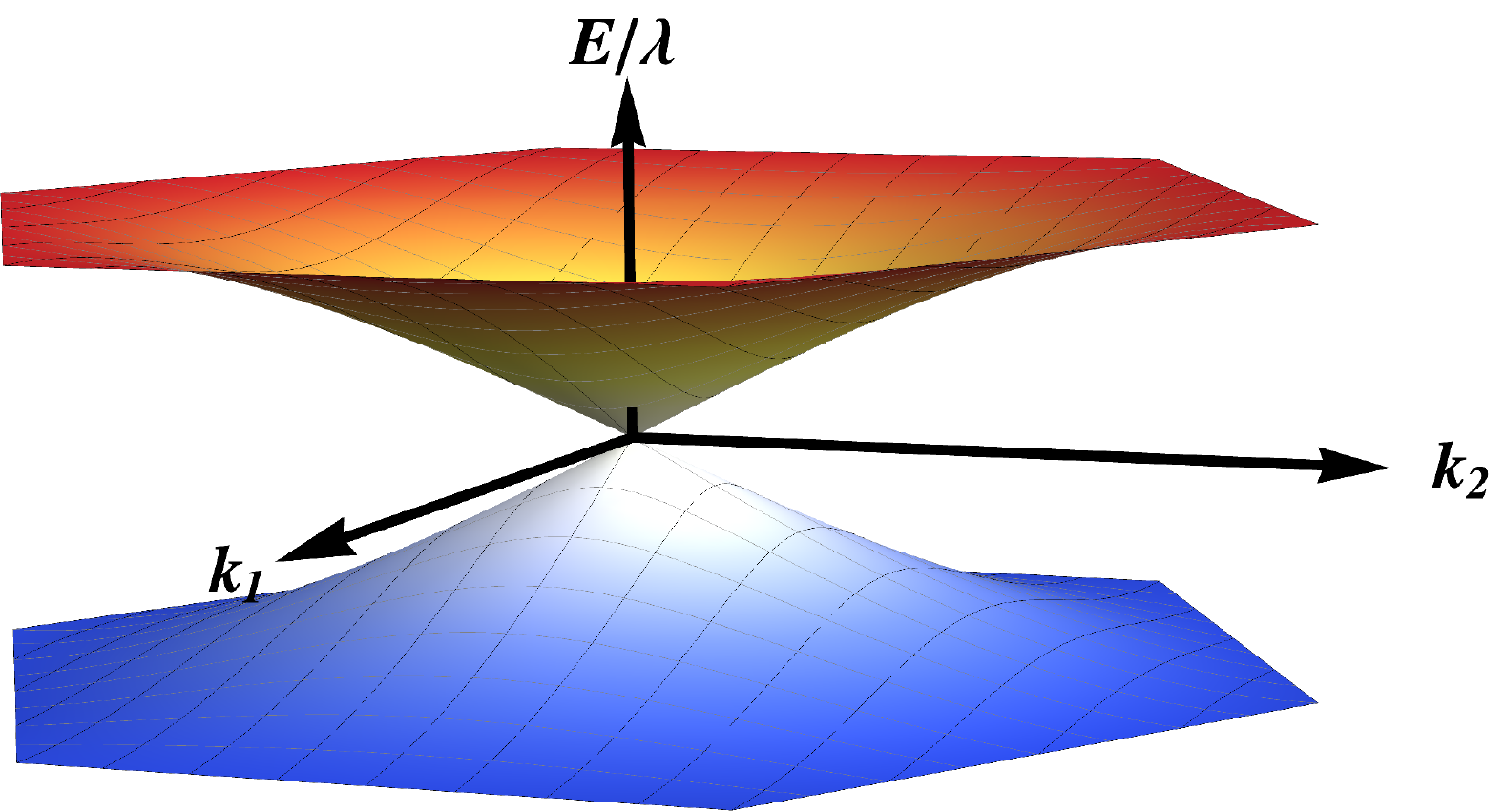}%
}
\hfill
\caption{Gapless behavior of Wilson loop Hamiltonian and surface-states of three-dimensional, octupolar topological insulator along $[1,1,1]$ direction. (a) The Wilson loop bands are two-fold degenerate and exhibit four-fold degeneracy only at the center of hexagonal, surface Brillouin zone. (b) The two-fold degenerate surface- conduction and valence bands meet at the Dirac point located at the center of surface Brillouin zone.
Here, $k_1$ and $k_2$ axes are perpendicular to the [111] direction, and we have set $(t_s, t_p, t_d) = (-\frac{\lambda}{\sqrt{3}}, \lambda, \frac{\lambda}{\sqrt{2}}) $ with $\lambda$ and $\Delta > 0$.  
}
\label{fig:(111)}
\end{figure}

For first-order TIs, $\Omega_3(k_1,k_2)$ vanishes at $\bar{\Gamma}: \; (k_1, k_2)=(0,0)$, and $\bar{M}: \; (k_1, k_2)=(0,0)$ points of the hexagonal, transverse or surface BZ. Furthermore, $W_3$ maps to $-\mathbbm{1}$ and $\mathbbm{1}$ at $\bar{\Gamma}$ and $\bar{M}$ points, respectively. This gives rise to the interpolation of WLH spectra ($\pm \lambda_3$) between $0$ and $2 \pi$. In contrast to this, $\Omega_3(k_1,k_2)$ for octupolar TIs only vanishes at $\bar{\Gamma}$. As $d$-wave terms gap out WLH spectra at $\bar{M}$ point, the WL of octupolar TI cannot interpolate between $\pm \mathbbm{1}$. 
The behavior of $\lambda_3(k_1,k_2)$ is shown in Fig.~\ref{fig:foti1}. Due to the four-fold degeneracy of WLH spectra at $\bar{\Gamma}$ point, we expect the surface-states for $(1 1 1)$ surface to be described by four-component, massless Dirac fermions. 
For the BBH model \emph{augmented by a finite $\Delta$}, the surface-states close to the zone center obtain an isotropic dispersion,
\begin{align}
\mathcal E_{(111)}^\pm(k_1, k_2) \approx \pm \frac{|t_p~ \Delta|}{3} \sqrt{ k_1^2 + k_2^2}.
\end{align}
With increasing deviations from the zone center, the Dirac cone acquires  a trigonal warping due to the  threefold rotational symmetry about the [1 1 1] axis [see Appendix~\ref{app:surface} for more details].
We note that for the BBH model of a three-dimensional 3rd order TI $\Delta = 0$, which would lead to flat surface states.
The solution of surface-states dispersion is illustrated in Fig.\ref{fig:hoti1} and the detailed form is presented in Appendix~\ref{app:surface}.
We further note that the spin-orbital texture of surface-states in the immediate vicinity of Dirac point can be reasonably approximated by linearized form of $\Omega_3$, which allows us to identify the surface Dirac point as $SO(5)$ vortex.

\section*{Beyond the BBH class}
The models of HOTIs discussed above are unitarily related to the BBH models, augmented by a general set of parameters, and maybe considered as models belonging to the `BBH class'.  
The distinguishing properties of these models under periodic boundary condition are (i) only a subset of TRIM locations support band inversion with respect to $\Gamma_{2D}$; (ii) only the body diagonals may be classified by $\pi_1(S^1)$.
By breaking the cubic symmetry, it is possible to construct models of HOTIs that do not satisfy these properties, and, thus, do not belong to the BBH class. 
An interesting class of models that lie beyond the BBH-class is constructed by replacing $f_d^l(\bs k)$ with $g_d^{i,j}(\bs l) = \sin{k_i} \sin{k_j}$ where $1 \leq i< j \leq D$.
At $D=2$, this corresponds to replacing $B_1$ harmonic $(\cos k_x - \cos k_y)$ in Eq.~\ref{eq:BBH-N} by $B_2$ harmonic $\sin k_x \sin k_y$ of $C_{4v}$ point group.
In $D$ dimensions, the vector field is given by 
\begin{align}
\bs N'(\bs k) =& [t_p f^1_p(\bs{k}),..,t_p f^D_p(\bs{k}), t_d g_d^{1,2}(\bs{k}),.., t_d g_d^{D-1,D}(\bs{k}), \nonumber \\ 
&  t_s f_s(\bs{k})], 
\label{eq:nonBBH-N}
\end{align}
and the Bloch Hamiltonian is $\hat{H}'(\bs k) = \sum_{j}^{M+1} N_j'(\bs k) \Gamma_j$, where $M =D(D+1)/2$.
Since there are $D(D-1)/2$  $d$-wave terms in \eqref{eq:nonBBH-N}, the coset space changes to $\frac{Spin(M + 1)}{Spin(M)}$, which readily distinguishes between the two classes of models at $D \geq 3$.  
Interestingly, akin to 1st order TIs, this non-BBH class of HOTIs support band inversion with respect to $\Gam_{M+1}$ at all TRIM locations, and admits Fu-Kane's $\mathbb{Z}_2$ index.
In contrast to 1st order TIs,  only the principal axes may be classified by $\pi_1(S^1)$, because along the diagonal axes $\hat H'(\bs k)$ is described by $3$-components vectors, and the corresponding unit vectors map these axes to $S^2$.
Such maps do not support a non-trivial 1st homotopy classification.
The Wilson lines along principal axes display gapless spectra and $\lambda_{x}(k_y)$ interpolates between $0$ and $2\pi$ (winding of Wannier charge centers).
Consequently, under the slab geometry along principal axes and the square geometry of Fig.~\ref{fig:sq} one finds gapless helical edge-modes, which connect bulk conduction and valence bands. 
\emph{However, the Wilson lines and the surface-states along $[1, \pm 1]$ directions display gapped spectra}. 
Therefore, under open boundary conditions with the diamond geometry of Fig.~\ref{fig:dia}, one obtains gapped edge modes and zero-energy, corner-localized, mid-gap states. 
The induced quadrupole moment will have $B_1$ symmetry. 
Here, we learn two important lessons:  (i) \emph{all high-symmetry lines joining band-inversion points are not required to support $\pi$ Berry phase or time-reversal polarization}; (ii) \emph{there exist HOTI models that display fully connected spectra of Wilson lines along principal axes}. 

Since the high-symmetry axes of the two classes of HOTIs have a complementary behavior, their mixing can produce HOTIs where no axis can support a non-trivial $\pi_1(S^1)$ classification. 
Consequently, the Wilson loop and edge/surface spectra along all high-symmetry directions are gapped.
Such models may be called ``ideal HOTI'' due to the absence of gapless surface states under any geometry. 
Generic planes lying in between the Dirac points in Kramers degenerate Dirac semimetals are two-dimensional examples of such HOTIs~\cite{tyner2020, szabo2020, wieder2020}, which admit the quantized non-Abelian Berry flux as a robust bulk invariant~\cite{tyner2020}.

\section*{Discussion}
In this work we formulated a unified, gauge-invariant description of bulk-topology through first-homotopy classification of high-symmetry lines and second homotopy classification of high-symmetry planes.
Our analytical results show that the Wilson loop spectra of HOTIs have strong direction dependence. The presence of band-inversion between the center and the corner of cubic Brillouin zone gives rise to gapless (gapped) spectra along body diagonals (principal axes). 
Consequently, $D$-th order TIs support gapless (gapped) surface-states along the body diagonal (principal axes).
In the class of models we considered here, the gapless surface-states take the form of $(D-1)$-dimensional Dirac fermions.
For similar reasons, gapless surface states are also present in $n$-th order TIs with $n \leq D$ under suitable sample geometries. 
\emph{Since recent experiments on engineered systems have simulated quadrupolar and octupolar topological insulators, our predictions for the gapless surface states and their relationship with corner-states can be directly verified with such experimental set up}.

Since topologically protected states at crystal terminations play a key role in determining the nature of  bulk-topology, corner or hinge localized states have been designated as the defining signatures of higher order topology.
Due to their sub-extensive nature, however, such states are usually not accessible to angle resolved photoemission spectroscopy, which is a powerful probe for elucidating band-topology of solid-state systems.
The Dirac-like surface-state we obtain here, therefore, offers an avenue for angle resolved photoemission spectroscopy, to directly address  higher order topology, if a suitable cleavage surface is available. 
Further, our results indicate the need for distinguishing between surface state signatures of higher-order and weak topology, as both may support Dirac cones on a subset of crystalline-symmetry preserving surface terminations. 

Our analysis can also be applied to other models of HOTIs, with stronger resemblance to 1st-order TIs. 
An interesting class of such models is obtained by using $^DC_2$ number of $\sin k_a \sin k_b$ type $d$-wave harmonics, which are odd under the mirror operation $k_a \to - k_a$ (or $k_b \to -k_b$).  
Akin to the 1st-order TI, the resulting model supports band inversion at all TRIM points, and admits Fu-Kane's $\mathbb{Z}_2$ index. 
The contrast between the two classes of HOTIs is succinctly  reflected by  the  localization patterns of  mid-gap states in crystalline-symmetric sample-geometries, characterized by the presence of surfaces perpendicular to \emph{all} high-symmetry axes, as exemplified by Fig.~\ref{fig:octa}.
Two-dimensional topological insulators with gapped Wilson loop and edge/surface spectra along all high-symmetry directions can be obtained by combining the  $f_d^{l}(\bs k)$ and $g_d^{i,j}(\bs k)$ type harmonics in Eqs.~\eqref{eq:BBH-N} and \eqref{eq:nonBBH-N}.
The resulted TIs generically lack the  $\mathbb{Z}_2$ particle-hole symmetry of BBH type models, and corner-states are found at finite energies under all crystalline-symmetry preserving sample geometries. 

Beside its influence over surface state properties, the 1D winding number  discussed here also controls the period of Bloch oscillations~\cite{li2016,holler2018,liberto2020}.
Thus, our 1D winding number based diagnostic of higher order topology is accessible to experiments, particularly for HOTIs realized in ultracold atoms setups. 
While direct experimental probes for the 2D and 3D winding numbers  discussed here are presently unavailable, these topological invariants can be ``measured'' in gedankenexperiments by inserting flux tubes and monopoles, respectively~\cite{tyner2020,tyner2022}. 
In particular, a non-zero quantized  flux (flux tunneling configuration)  in the Brillouin zone would result in isolated zero modes being trapped at the vortex (monopole) core.
The number of such zero-modes reveals the magnitude of the 2D or 3D winding numbers.

\acknowledgements{This work was supported by the National Science Foundation MRSEC program (DMR-1720139) at the Materials Research Center of Northwestern University. 
The work of S.S. at Rice University was supported by the U.S. Department of Energy Computational Materials Sciences (CMS) program under Award Number DE-SC0020177.
S.S. would like to thank Marco di Liberto, Giandomenico Palumbo, and Ming Yi for helpful discussions.
The plots in Fig.~\ref{fig:octagon} were produced by the \emph{Pybinding} package~\cite{pybind}.}




\bibliographystyle{plain}


\appendix
\onecolumngrid
\clearpage


\section{Explicit expressions of $\omega_{ab}$'s} \label{app:Om}
\begin{align}
& \om_{12} =  t_p^2 \sin{(\sqrt{2}k_1)}; 
\; 
\om_{13} = t_p^2 \sin{\qty(\frac{k_1}{\sqrt{2}} + \frac{\sqrt{3} k_2}{\sqrt{2}})}; 
\nn \\
& \om_{23} = - t_p^2 \sin{\qty(\frac{k_1}{\sqrt{2}} - \frac{\sqrt{3} k_2}{\sqrt{2}})};
\;  
\om_{14} = - \om_{24} =  2 t_p t_d \sin^2{\frac{k_1}{\sqrt{2}}} 
\nn \\
& \om_{34} = - 2 t_p t_d \sin{\frac{k_1}{\sqrt{2}}}  \sin{\frac{\sqrt{3} k_2}{\sqrt{2}}};
\;
\om_{35} = - \frac{2t_p t_d}{\sqrt{3}} \qty[1 - \cos{\frac{k_1}{\sqrt{2}}} \cos{\frac{\sqrt{3} k_2}{\sqrt{2}}} ]; 
\nn \\
& \om_{15} = \frac{t_p t_d}{\sqrt{3}} \qty[1 + \cos{(\sqrt{2} k_1)} - 2 \cos{\qty(\frac{k_1}{\sqrt{2}} + \frac{\sqrt{3} k_2}{\sqrt{2}})} ] ; \nn \\
& \om_{25} = \frac{t_p t_d}{\sqrt{3}} \qty[1 + \cos{(\sqrt{2}k_1)} - 2 \cos{\qty(\frac{k_1}{\sqrt{2}} - \frac{\sqrt{3} k_2}{\sqrt{2}})} ]; \nn \\
& \om_{45} = \frac{4 t_d^2}{\sqrt{3}} \sin{\frac{k_1}{\sqrt{2}}} \qty(\cos{\frac{k_1}{\sqrt{2}} } - \cos{\frac{\sqrt{3} k_2}{\sqrt{2}}}
).
\end{align}


\section{Gapless surface-states of octupolar model} \label{app:surface}
Here, we note the key intermediate steps for the derivation of the [111] surface states.
In the $(k_1, k_2, k_3)$ coordinates the Hamiltonian takes the form,
\begin{align}
H_{D=3} = \bs N \cdot \bs \Gam',
\label{eq:ham111}
\end{align}
where $\bs \Gam'$ are a set of 6  mutually anti-commuting $8 \times 8$ matrices, and
\begin{align}
& \frac{N_1}{t_p} = \cos \frac{k_3}{\sqrt{3}} \left[\sin \frac{k_1}{\sqrt{2}}  \cos \frac{k_2}{\sqrt{6}} +\sin \frac{k_2}{\sqrt{6}}  \cos \frac{k_1}{\sqrt{2}} \right] 
+\sin \frac{k_3}{\sqrt{3}}  \left[\cos \frac{k_1}{\sqrt{2}}  \cos \frac{k_2}{\sqrt{6}} -\sin \frac{k_1}{\sqrt{2}}  \sin \frac{k_2}{\sqrt{6}} \right] \nn \\
& \frac{N_2}{t_p} = \cos \frac{k_3}{\sqrt{3}}  \left[\sin \frac{k_2}{\sqrt{6}}  \cos \frac{k_1}{\sqrt{2}} -\sin \frac{k_1}{\sqrt{2}}  \cos \frac{k_2}{\sqrt{6}} \right]
+ \sin \frac{k_3}{\sqrt{3}}  \left[\sin \frac{k_1}{\sqrt{2}}  \sin \frac{k_2}{\sqrt{6}} +\cos \frac{k_1}{\sqrt{2}}  \cos \frac{k_2}{\sqrt{6}} \right] \nn \\
& \frac{N_3}{t_p} = 	\sin \frac{k_3}{\sqrt{3}} \cos \left(\sqrt{\frac{2}{3}} k_2\right) -\sin \left(\sqrt{\frac{2}{3}} k_2\right) \cos \frac{k_3}{\sqrt{3}}  \nn \\
& \frac{N_4}{2t_d} =   \sin \frac{k_1}{\sqrt{2}}  \sin \frac{k_3}{\sqrt{3}} \cos \frac{k_2}{\sqrt{6}} 
+  \sin \frac{k_1}{\sqrt{2}}  \sin \frac{k_2}{\sqrt{6}}  \cos \frac{k_3}{\sqrt{3}} \nn \\
& \frac{\sqrt{3} N_5}{2 t_d} =  \cos \frac{k_3}{\sqrt{3}} \left[ \cos \left(\sqrt{\frac{2}{3}} k_2\right)-  \cos \frac{k_1}{\sqrt{2}}  \cos \frac{k_2}{\sqrt{6}}  \right]
+ \sin \frac{k_3}{\sqrt{3}} \left[  \sin \left(\sqrt{\frac{2}{3}} k_2\right) +  \sin \frac{k_2}{\sqrt{6}}  \cos \frac{k_1}{\sqrt{2}}  \right] \nn \\
& \frac{N_6}{t_s} = \cos \frac{k_3}{\sqrt{3}} \left[-2 \cos \frac{k_1}{\sqrt{2}}  \cos \frac{k_2}{\sqrt{6}} -\cos \left(\sqrt{\frac{2}{3}} k_2\right)\right]
+ \sin \frac{k_3}{\sqrt{3}}  \left[2 \sin \frac{k_2}{\sqrt{6}}  \cos \frac{k_1}{\sqrt{2}} -\sin \left(\sqrt{\frac{2}{3}} k_2\right)\right]
+ \Delta  .
\end{align}
After a sequence of unitary transformations one can solve for the states on the (111) surface for the system occupying the half space $r_3 > 0$ (or, equivalently, $r_3 < 0$), where $r_3$ is the position space conjugate of $k_3$.
We find a pair of twofold degenerate bands, described by
\begin{align}
\mc E_{(111)}(k_1, k_2) = \pm \qty|t_s \Delta| ~\sqrt
{(\cos{\theta_{3}} \sin{\theta_{5}} \sin{\phi_{1}})^2 
+ ( \cos{\theta_{3}} \sin{\theta_{5}} \cos{\phi_{1}} \cos{\phi_{2}}
+ \sin{\theta_{3}} \sin{\phi_{2}}
)^2},
\end{align}
where
\begin{align}
& \theta_{1} = \tan^{-1} \frac{\sin\qty(\frac{k_1}{\sqrt{2}} + \frac{k_2}{\sqrt{6}})}{\sin\qty(\frac{k_1}{\sqrt{2}} - \frac{k_2}{\sqrt{6}})}; 
\qquad
\theta_{2} = - \tan^{-1} \frac{t_p \cos\frac{k_2}{\sqrt{6}}}{t_d \sin\frac{k_1}{\sqrt{2}}}; 
\qquad
\theta_{3} = \tan^{-1} \frac{2 t_d \qty[
\cos\qty(\frac{2}{\sqrt{6}} k_2 ) - \cos{\frac{k_1}{\sqrt{2}}} \cos{\frac{k_2}{\sqrt{6}}}
]}
{\sqrt{3} t_s \qty[
\cos\qty(\frac{2}{\sqrt{6}} k_2 ) + 2 \cos{\frac{k_1}{\sqrt{2}}} \cos{\frac{k_2}{\sqrt{6}}}
]}; \nn \\
& f_{1} = \sqrt{
\sin^2\qty(\frac{k_1}{\sqrt{2}} + \frac{k_2}{\sqrt{6}}) + \sin^2\qty(\frac{k_1}{\sqrt{2}} - \frac{k_2}{\sqrt{6}})
}; \qquad
f_{2} = \sqrt{
\qty(t_p \cos\frac{k_2}{\sqrt{6}})^2 + \qty(t_d \sin\frac{k_1}{\sqrt{2}})^2 
}; \nn \\
& f_{3} = \sqrt{
(2 t_d)^2 \qty[
\cos\qty(\frac{2}{\sqrt{6}} k_2 ) - \cos{\frac{k_1}{\sqrt{2}}} \cos{\frac{k_2}{\sqrt{6}}}
]^2 
+ (\sqrt{3} t_s)^2 \qty[
\cos\qty(\frac{2}{\sqrt{6}} k_2 ) + 2 \cos{\frac{k_1}{\sqrt{2}}} \cos{\frac{k_2}{\sqrt{6}}}
]^2
}; \nn \\
& \theta_{4} = \tan^{-1} \frac{t_p f_{1}}{f_{2}}; \qquad
\theta_{5} = \tan^{-1} \frac{\sqrt{t_p^2 f_{1}^2 + f_{2}^2}}{f_{3}}; \nn \\
& \tilde g_1 = \cos\theta_{1} g_1 + \sin\theta_{1} g_2;
\qquad
\tilde g_2 = -\cos\theta_{4} \qty( \sin\theta_{1} g_1 - \cos{\theta_{1}} g_2 ) 
- \sin\theta_{4} \qty( \sin\theta_{2} g_3 + \cos{\theta_{2}} g_4 ); \nn \\
&\tilde  g_3 = \cos\theta_{2} g_3 - \sin\theta_{2} g_4; \qquad
\tilde g_5 = \cos\theta_{3} g_5 + \sin{\theta_{3}} g_6 \nn \\
& \tilde g_4 = \cos{\theta_{5}} \qty[
\sin{\theta_{4}} (\cos{\theta_{1}} g_2 - \sin{\theta_{1}} g_1 )
+ \cos{\theta_{4}} (\sin{\theta_{2}} g_3 + \cos{\theta_{2}} g_4 )
]
+ \sin{\theta_{5}} \qty[
\sin{\theta_{3}} g_5 - \cos{\theta_{3}} g_6
] \nn \\
& \phi_{1} = \tan^{-1}\frac{\sqrt{\tilde g_1^2 + \tilde g_2^2 + \tilde g_3^2}}{\tilde g_4}; \qquad
\phi_{2} = \tan^{-1}\frac{\sqrt{\tilde g_1^2 + \tilde g_2^2 + \tilde g_3^2 + \tilde g_4^2}}{\tilde g_5},
\end{align}
with $g_j$ being the coefficient of $\Gamma_j'$ multiplying $\sin{\frac{k_3}{\sqrt{3}}}$ in \eq{eq:ham111}.
The conduction and valence bands are two-fold degenerate and display four-fold degeneracy only at the center of heaxgonal surface BZ.
Close to the zone center, in terms of $\{k_\perp, \phi\} = \{\sqrt{k_1^2 + k_2^2}, \arctan(k_2/k_1)\}$,  
\begin{align}
&\frac{\mc E_{(111)}(k_\perp, \phi)}{|t_s \Delta|} = \frac{|t_p|}{3|t_s|} k_\perp  
+ \frac{3 (t_d^2  + 2 t_p^2) t_s^2 - 4 t_{p}^4}{216 |t_{p}| |t_{s}|^3 } k_\perp^3 \nn \\
&+ \frac{12 t_s^2 t_p^4 \left(t_s^2 (8 \cos (6 \phi )+5)- 10 t_d^2 \right)-60 t_d^2 t_p^2 t_s^4 (5 \cos (6 \phi )-7)+15 t_d^4 t_s^4 (16 \cos (6 \phi )-17)-240 t_p^6 t_s^2+80 t_p^8}{51840 |t_p|^3 |t_s|^5} k_\perp^5 
 + o(k_\perp^7),
\end{align} 
which leads to the asymptotic form of the dispersion noted in the main section of the paper.
We note that the warping term arises at $o(k_\perp^5)$, and it is present at $t_d = 0$, i.e. for 1st order TIs.

\end{document}